\documentclass[twocolumn,showpacs,preprintnumbers,amsmath,amssymb]{revtex4}

\usepackage{graphicx}
\usepackage{dcolumn}
\usepackage{bm}

\begin{document}

\title{{\it In situ} photoemission study of the room-temperature
ferromagnet ZnGeP$_2$:Mn}

\author{Y.~Ishida$^1$, D. D.~Sarma$^2$, K.~Okazaki$^1$, J.~Okabayashi$^1$, 
J. I.~Hwang$^1$, H.~Ott$^3$, A.~Fujimori$^1$,
G. A.~Medvedkin$^4$, T.~Ishibashi$^4$, 
K.~Sato$^4$}

\affiliation{$^1$Department of Physics, University of Tokyo,
Bunkyo-ku Tokyo 113-0033, Japan} 

\affiliation{$^2$Solid State and Structural Chemistry Unit,
Indian Institute of Science, Bangalore 560 012 India} 

\affiliation{$^3$Institut f\"{u}r Experimentalphysik 
Freie Universit\"{a}t Berlin, 
Arnimallee 14 14195 Berlin, Germany}

\affiliation{$^4$Tokyo University of Agriculture and Technology,
Koganei Tokyo 184-8588, Japan}


\date{\today}

\begin{abstract}
The chemical states of the ZnGeP$_{2}$:Mn interface, which shows
ferromagnetism above room-temperature, has been studied by photoemission
spectroscopy. Mn deposition on the ZnGeP$_2$ substrate heated to 
400$^{\circ}$C induced 
Mn substitution for Zn and then the formation of metallic Mn-Ge-P 
compounds. Depth profile studies have shown that Mn 3$d$ electrons 
changed their character from itinerant to localized along the depth, 
and in the deep region, dilute divalent Mn species 
($\textless$ 5 \% Mn) 
was observed with a coexisting metallic Fermi edge of 
non-Mn 3$d$ character. The possibility of hole doping 
through Mn substitution for Ge and/or Zn vacancy is discussed. 
\end{abstract}



\maketitle
The successful synthesis of the III-V-based ferromagnetic diluted magnetic 
semiconductors (DMSs) has opened up a large number of 
exciting functionalities such as non-volatile memories, spin injection, 
and the optical control of ferromagnetism in 
semiconductor devices \cite{Ohno}. 
However, the reported Curie temperature ($T_C$) to date 
has been limited to 120 K realized in 
Ga$_{1-x}$Mn$_x$As, due to the limited ability of incorporating Mn ions and 
{\it p}-type carriers 
in the III-V-based DMS. Overcoming this limitation 
and increasing the $T_C$ hopefully above the room temperature (RT) has 
been a challenging subject. 

Recently, there have been several reports on high-$T_C$ 
ferromagnetism in new DMSs 
\cite{RTDMSs}. 
Among them are intriguing reports 
by Medvedkin {\it et al}.\ on 
Mn incorporated II-IV-V$_{2}$ chalcopyrite 
semiconductors CdGeP$_{2}$ \cite{CGP} and 
ZnGeP$_{2}$ \cite{ZGP}, which show ferromagnetism 
above RT. A high concentration of Mn ions is incorporated in 
the semiconductors through deposition 
of Mn metal and annealing.
The use of II-IV-V$_2$ semiconductor as a host material is 
attracting, since, while it is an average III-V semiconductor, 
it has two types of cation sites.
This distinct property compared to the binary III-V semiconductors
allows us to functionalize each cation site in different ways, 
e.g., magnetic ion doping 
at the II site and acceptor doping at the IV site. 
In addition, the chalcopyrite semiconductors have many natural defects, 
e.g., zinc ion on the germanium site (Zn$_{\rm Ge}$), zinc 
vacancy (V$_{\rm Zn}$), etc, in ZnGeP$_2$, which may 
also act as acceptors \cite{defect}. 
This rich chemistry stimulated many theoretical calculations on 
II-IV-V$_2$:Mn \cite{Calc}. 
However, there has been little experimental information about 
the chemical reactions and chemical products in this system.

Photoemission spectroscopy (PES) is a powerful tool to investigate 
chemical reactions in surface and interfacial regions. 
Particularly for studying the 
Mn-doped chalcopyrites, where the Mn density varies 
as a function of the depth, 
PES combined with ion sputtering provides 
us with the depth profile of the interface. At the same time, PES 
provides insight into the electronic structures of DMSs \cite{PES}, 
particularly into the Mn 3$d$ partial density 
of states (PDOS) through resonant PES and through subsequent analyses of 
the $p$-$d$ exchange interaction 
between the Mn $3d$ electrons and the host semiconductors. 

Ultraviolet photoemission (UPS) and x-ray photoemission (XPS) measurements 
were performed at BL-18A of Photon Factory. 
The total energy resolution 
of the VG CLAM analyzer 
including temperature broadening was 800 meV 
for XPS and 200 meV for UPS. 
Mn metal (99.999 \%) was evaporated {\itshape in situ\/} 
onto the (001) surface of single crystal ZnGeP$_{2}$ at 400$^\circ$C 
\cite{ZGP}. The deposition rate was 3 \AA/min, determined by a 
quartz thickness monitor. After the Mn deposition, the sample was 
post-annealed at 400$^{\circ}$C for 5 min and then cooled to RT. 
All the spectra were taken at RT.
Ar$^+$-ion sputtering 
at 1.5 kV
was used both for cleaning and etching. 
Sputter-etching rate was $\sim$ 2 {\AA}/min. Surface cleanliness 
was checked by core-level XPS. The pressure was below 
7 $\times$ 10$^{-10}$ Torr during the measurements. 
The magnetization was measured
{\it ex situ} using a 
SQUID magnetometer. 
(Quantum Design MPMS). 
For comparison, polycrystalline MnP ($T$$_C$ = 290 K \cite{MnP}) was 
also measured. 

\begin{figure}[htb]
\begin{center}
\includegraphics[width=8.0cm]{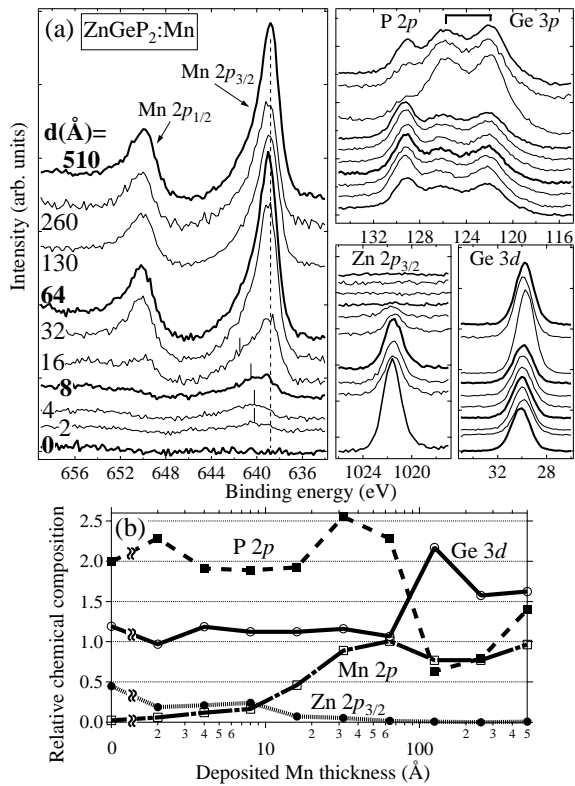}
\caption{\label{fig1}Core-level spectra of 
ZnGeP$_2$:Mn for various Mn thickness deposited at 
400$^{\circ}$C. (a) Raw spectra. The vertical scale is counts per second. 
0 \AA\ corresponds to the sputtered and annealed surface 
of the ZnGeP$_2$ substrate. (b) Core-level 
intensities as functions of deposited Mn thickness. For 
normalization, see text.}
\end{center}
\end{figure}

First, the effect of annealing 
and sputtering on the ZnGeP$_2$ substrate was studied. The 
core-level intensities above 300$^{\circ}$C showed strong Zn 
deficiency and weak Ge excess. [See the 
relative chemical composition at $d$ = 
0 \AA\ in Fig.\ \ref{fig1}(b), where 
the deviation from the stoichiometry of ZnGeP$_2$ is seen 
for the surface once annealed at 400$^{\circ}$C.] 
At the reported 
``optimum'' substrate temperature 400$^{\circ}$C, 
therefore, deposited Mn atoms may easily diffuse into 
the Zn-deficient surface region. 
This Zn-deficient region could be removed by 3-min Ar$^+$-ion sputtering 
(corresponding to $\sim$ 6 {\AA} thickness). Further sputtering did not 
change the core-level intensities of the substrate within the accuracy 
of 3 \%. This suggests that selective sputtering of different atomic species 
can be ignored, and we assumed that the as-sputtered substrate spectra 
rrepresent the bulk stoichiometry of ZnGeP$_2$. 

Next we show a series of spectra for Mn deposition on 
the 400$^{\circ}$C-annealed substrate. 
Each time after having taken a set of spectra for one Mn coverage, 
the deposited Mn was completely removed by prolonged sputtering 
and then Mn was newly deposited. 
Figure \ref{fig1}(a) shows the core-level spectra of the Mn-deposited surface, 
and Fig.\ \ref{fig1}(b) shows their intensities normalized to those for the 
as-sputtered ZnGeP$_2$. The Mn 2$p$ intensity has been normalized 
using the Mn 2$p$ and 
P 2$p$ intensity ratio of MnP. In the region of Mn thickness 
$d$ $<$ 64 \AA, one can see a 
monotonic increase and decrease of the Mn and Zn signals, respectively, 
without significant changes in the Ge and P intensities. This behaviour 
suggests that 
Mn atoms primarily substituted for Zn and a Zn$_{1-x}$Mn$_x$GeP$_2$-like 
compound was formed. 
In the Mn 2$p$$_{3/2}$ spectra below $d$ = 16 \AA, one can see some signal on 
the higher binding energy side [indicated by vertical bars 
in Fig.\ \ref{fig1}(a)] of the dominant metallic peak ($E_B$ = 638.7 eV, 
broken line). 
This indicates that a portion of 
the incorporated Mn atoms 
chemically reacted with the substrate and presumably became Mn$^{2+}$. 
In going from $d$ = 64 to 130 \AA, 
the Ge and P intensities suddenly changed. 
The saturation of the Mn intensity with sizable signals of Ge and P above 
$d$ = 130 \AA\ is a signature of atomic diffusion associated with the 
chemical reaction for such a thick Mn overlayer. 
The surface region above $d$ = 130 \AA\ consisted of Ge-rich, 
ternary metallic compound(s) 
of Mn, Ge, and P with possible inhomogeneity and/or phase mixture.
\begin{figure}[htb]
\begin{center}
\includegraphics[width=8.0cm]{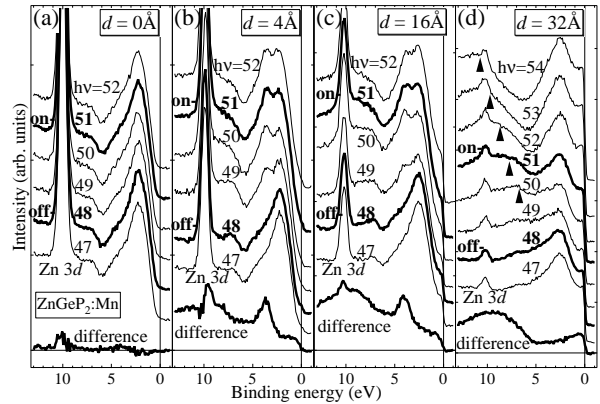}
\caption{\label{fig2}Valence-band spectra of ZnGeP$_2$:Mn in the 
3$p$-3$d$ core excitation region. On- and off-resonance energies are 
51 eV and 48 eV, respectively, and the spectra at the bottom show 
their difference spectra. Triangles in (d) indicate the constant 
kinetic energy of the Mn {\it M$_{3}$L$_{4,5}$L$_{4,5}$} Auger signal.}
\end{center}
\end{figure}

The valence-band spectra taken for photon energies 
in the Mn 3$p$-3$d$ core-excitation region 
are shown in Fig. \ref{fig2}. The spectra have been primarily normalized 
to the post-focusing Au mirror current and then to the intensities 
of the $d$ = 0 \AA\ spectra at the corresponding photon energies. 
The Mn 3$d$-drived spectra were obtained by 
subtraction between these normalized on- (51 eV) and 
off-resonance (48 eV) spectra. 
In this way, we could virtually eliminate 
the effect of the photon energy dependence of the host 
valence band. For $d$ = 4 and 16 \AA, one can see a peak 
located $\sim$ 4 eV below $E$$_F$
in the difference spectra. This observation is similar 
to the previous results of the Mn incorporated II-VI- and III-V-based 
DMSs \cite{PES} 
and thus is attributed to the localized nature of Mn 3$d$ electrons. 
However, above $d$ = 32 \AA, there was a change in the decay process 
of the Mn 3$p$ core hole and strong Mn 
{\it M$_{3}$L$_{45}$L$_{45}$} Auger peak 
replaces the $\sim$ 4 eV peak, indicating that the Mn 3$d$ electrons became 
itinerant. The disappearance of the $\sim$ 4 eV peak 
is in accordance with the Mn 2$p$ core-level 
spectrum [Fig. \ref{fig1}(a)], where the divalent Mn signal disappeared 
above $d$ = 32 \AA\ and the highly asymmetric line shape characteristic of 
a metallic Mn compound appeared. 
We note that MnP also shows Mn {\it M$_{3}$L$_{45}$L$_{45}$} 
Auger signals in the 
valence-band spectra \cite{Kakizaki}. 
However, the Mn $2p_{3/2}$ core-level peak 
of MnP was observed at $E_B$ = 639.2 eV, different from the peak positions 
observed above $d$ = 32 \AA\ in Fig.\ \ref{fig1}(a). 
As for the $d$ = 16 \AA\ spectrum, both a clear Fermi edge 
and the $\sim$ 4 eV peak is observed. 
The difference spectrum for 
$d$ = 16 \AA\ is almost a superposition of the $d$ = 4 \AA\ and 32 \AA\ 
spectra, as in the case of the Mn 2$p$ core-level spectra 
[Fig. \ref{fig1}(a)]. 
\begin{figure}[htb]
\begin{center}
\includegraphics[width=8.0cm]{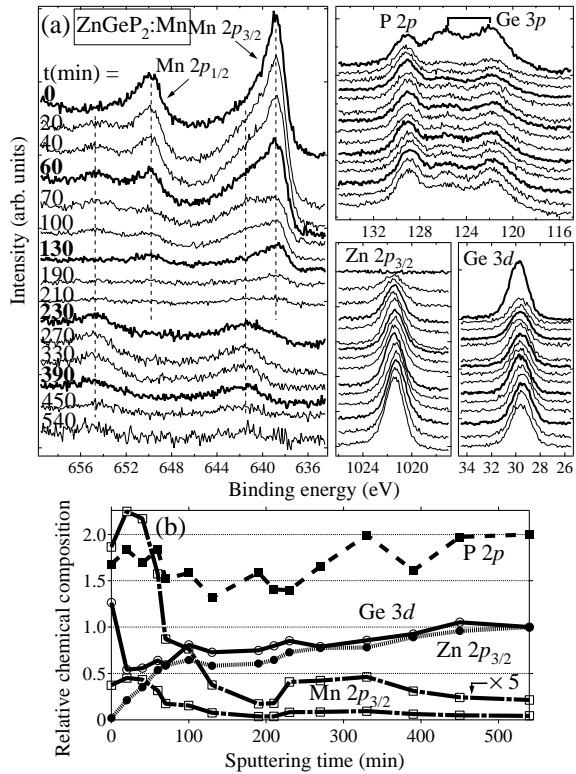}
\caption{\label{fig3}Core-level spectra of ZnGeP$_2$:Mn in the 
depth profile. (a) Raw spectra. The vertical scale is counts per second. 
(b) Core-level intensities as functions to sputtering time.}
\end{center}
\end{figure}

In order to study the chemical states formed underneath 
the surface metallic compound(s), ZnGeP$_2$:Mn 
of the nominal Mn thickness of 150 {\AA} was repeatedly sputter-etched 
without annealing and studied by PES. 
Figure \ref{fig3} (a) shows core-level spectra taken in the sputter-etching 
series and Fig.\ \ref{fig3} (b) shows their intensities [the same 
normalization as in Fig.\ \ref{fig1} (b)] as functions of sputtering time. 
The change in the first 20 min sputtering is attributed to the removal of the 
metallic Mn-Ge-P layer in the surface region. 
Subsequently, 
the Mn 2$p$$_{3/2}$ core level started to show a shoulder structure at 
$E$$_{B}$\ = 641.7 eV due to ionic Mn (Mn$^{2+}$ most likely). 
The systematic increase of this shoulder 
and the decrease of the metallic main peak between 20 to 70 
min sputtering indicate that these signals are 
originated from chemically different Mn species. 
After 100 min sputtering, the relative compositions became 
Zn:Ge:P $\sim$ 1:1:2, suggesting that the chalcopyrite-type matrix of 
Zn, Ge, and P plus dilute Mn was exposed [Fig.\ \ref{fig3} (b)]. 
After 230 min sputtering, the Mn signal became totally that of the 
ionic one [Fig.\ \ref{fig3} (a)], indicating that 
a dilute-Mn phase ($\textless$ 5 \% Mn) appeared. 
Since Zn:Ge:P $\sim$ 1:1:2 and 
the amount of Mn was small, 
it was difficult to determine from Fig.\ \ref{fig3}(b) which element 
Mn substituted for. Now, since the peak position of the ionic 
Mn 2$p$ signal below $\sim$ 200 min sputtering (in the intermediate phase) 
corresponds to that above $\sim$ 200 min sputtering (dilute Mn phase), 
the ionic Mn species in 
the former phase region may be precursors of those in the latter phase. 
\begin{figure}[htb]
\begin{center}
\includegraphics[width=8.0cm]{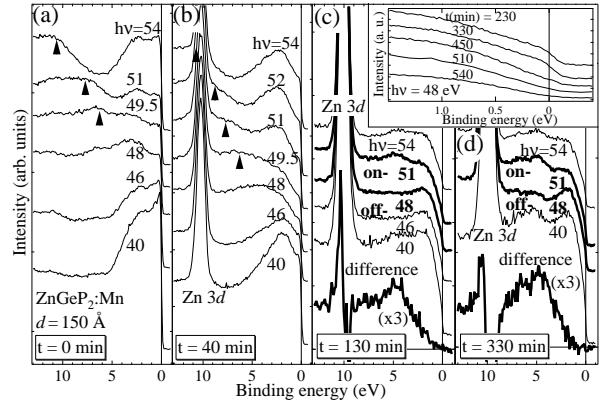}
\caption{\label{fig4}Valence-band spectra of ZnGeP$_2$:Mn ($d$ = 150 \AA) 
in the sputter-etching series. Triangles in (a) and (b) denote the 
{\it M$_{3}$L$_{4,5}$L$_{4,5}$} Auger signals. Inset shows the valence-band 
spectra near $E_F$ for sputtering time longer than 230 min.}
\end{center}
\end{figure}

Figure \ref{fig4} shows the valence-band spectra in the sputter-etching 
series for photon energies in 
the Mn 3$p$-3$d$ core excitation region. 
Before 80 min sputtering, the {\it M$_{3}$L$_{4,5}$L$_{4,5}$} Auger process 
was dominant 
[Fig.\ \ref{fig4} (a) and (b)] and after that resonant photoemission became 
dominant [Fig.\ \ref{fig4} (c) and (d)]. This indicates again that the Mn 3$d$ 
states changed their character from itinerant to localized along 
the depth profile. A peak at $E$$_{B}$\ $\sim$ 4 eV 
was seen after 80 min sputtering, and hence Mn was divalent 
in the deep region. 
The divalent Mn signal in the recent EPR measurements may come 
from this region \cite{EPR}. 
The Mn 3$d$-derived intensity near $E$$_F$ 
in the difference spectra is weak compared to that in 
Fig.\ \ref{fig2} (b). This indicates that the Mn 3$d$ states 
are more strongly localized in the deep bulk region 
than the Zn$_{1-x}$Mn$_x$GeP$_2$-like phase in the early stage of 
Mn deposition. Inset of Fig.\ \ref{fig4} shows 
valence-band spectra near $E_F$ after 230 min sputtering, where Mn had 
fully reacted with the substrate [Fig.\ \ref{fig3} (a)]. They clearly show a 
Fermi edge. Since the Mn 3$d$ PDOS is supressed near $E_F$, this Fermi edge 
comes from the valence band of the host semiconductor which was somehow doped 
with metallic charge carriers. 
Since isovalent substitution of Mn$^{2+}$ for Zn$^{2+}$ cannot produce 
carriers, Mn$^{2+}$ may have substituted for the Ge site 
and/or Mn incorporation simultaneously 
induced defects such as, e.g.,\ V$_{\rm Zn}$ and Zn$_{\rm Ge}$, all of which 
produce hole carriers. The Fermi edge became 
obscure after 540 min sputtering, in accordance with the diminishing 
Mn 2$p$ core-level intensity (Fig.\ \ref{fig3}). 
In the 
sputter-etching series, too, no MnP signal was observed. 
\begin{figure}[htb]
\begin{center}
\includegraphics[width=8.0cm]{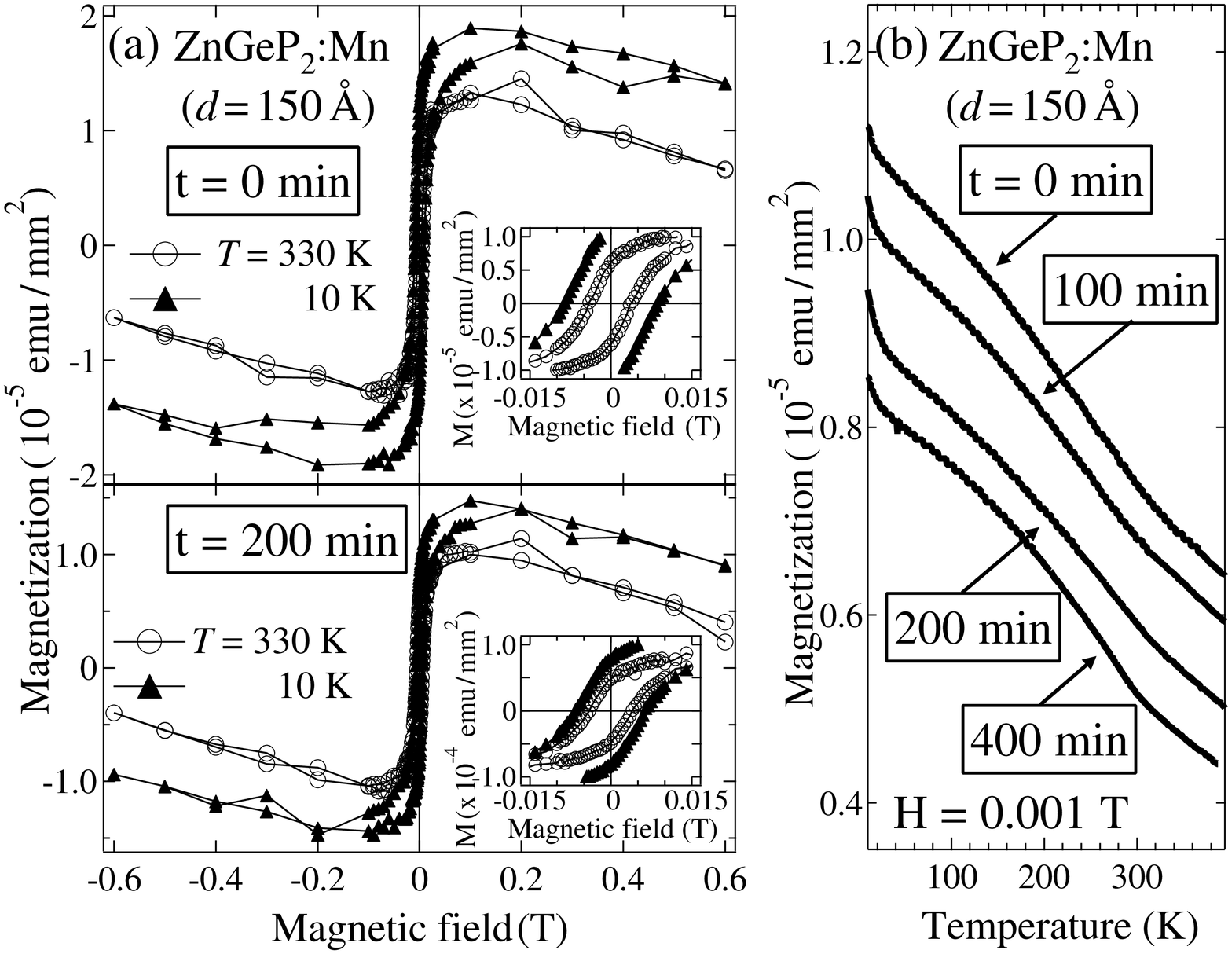}
\caption{\label{fig5}Magnetization per sample area of ZnGeP$_2$:Mn 
($d$ = 150 {\AA}) prepared at 400$^\circ$C. 
(a) Magnetization curves at $T$ = 10 K and 330 K of as-prepared 
ZnGeP$_2$:Mn (upper) 
and those after having removed the surface 
metallic Mn compound (lower). From these data, magnetization per Mn atom 
is crudely estimated to be 1.5 $\mu$$_B$. Insets 
show the magnified plot revealing hysterisis. (b) Magnetization 
versus temperature in the sputtering series.}
\end{center}
\end{figure}

We have also studied the depth profile of ZnGeP$_2$:Mn of 
the 200 {\AA} nominal thickness Mn annealed at 200$^{\circ}$C. 
Up to 800 min sputtering, Mn signal was observed together with Zn, Ge, 
and P signals, indicating that diffusive reaction of Mn into the 
substrate occurred already at 200$^{\circ}$C. 
However, the Mn 2$p$$_{3/2}$ signal always appeared at 
$E$$_{B}$\ = 638.7 eV without any divalent signal. 
Correspondingly, only the Mn 
{\it M$_{3}$L$_{4,5}$L$_{4,5}$} Auger peak was observed 
in the valence-band spectra. 
We therefore conclude that the annealing at 200$^{\circ}$C was 
insufficient for the Mn atoms to be chemically incorporated 
in the host semiconductor 
as divalent ions with localized nature of Mn 3$d$ electrons. 

Figure \ref{fig5}(a) shows the magnetization of as-prepared 
ZnGeP$_2$:Mn ($d$ = 150 {\AA} at 400$^{\circ}$C) and that 
after having removed the surface metallic Mn compound 
(200 min sputtered). Clear hysterisis was observed between 
10 K and 330 K for the 150 {\AA} deposited sample. 
Surprisingly, one can observe nearly the same magnetization 
at both 330 K and 10 K even after removing 
the surface metallic Mn compound, 
indicating that the divalent Mn phase 
deep in the substrate ($\textless$ 5 \% Mn) was probably a RT ferromagnet. 
Figure \ref{fig5}(b) clearly shows 
ferromagnetism up to $\sim$ 400 K. One can see 
a kink at $\sim$ 300 K which may correspond to the $T_C$ = 312 K of bulk 
Zn$_{1-x}$Mn$_x$GeP$_2$ \cite{Cho}. 
Another anomaly at 20 - 50 K may also correspond to 
the 47 K anomaly of bulk Zn$_{1-x}$Mn$_x$GeP$_2$. 
Therefore, the present sample may contain 
Zn$_{1-x}$Mn$_x$GeP$_2$ as a minority component. Recently, 
ZnGeP$_2$:Mn prepared at 550$^{\circ}$C was studied, and the 
$M$($T$) curve showed pronounced singularities at 
$\sim$ 318 K and 20 - 50 K and small magnetization tailing 
above 318 K, the behaviour of which is more similar to 
bulk Zn$_{1-x}$Mn$_x$GeP$_2$ than the present sample \cite{550C}. 
We note that the bulk Zn$_{1-x}$Mn$_x$GeP$_2$ was 
reported to be electrically insulating, while 
the present sample showed a metallic Fermi edge. Further 
studies are necessary, in particular to understand 
the relationship between the 
magnetic behaviour, the carrier density, and the preparation temperature. 

In conclusion, we have observed spectral features of localized, most likely 
divalent Mn 3$d$ states incorporated into the host ZnGeP$_2$ in 
the thin Mn-deposited surface region (probably as a 
Zn$_{1-x}$Mn$_x$GeP$_2$) and in deep region 
below the surface metallic Mn compound after thick Mn deposition. 
A Fermi edge was observed in the 
deep region, indicating that carrier doping took place 
in that region of ZnGeP$_2$:Mn. RT ferromagnetism 
was observed after removing the 
surface Mn compound. 
This indicates that ferromagnetism in ZnGeP$_2$:Mn is caused by 
the dilute Mn ions in the deep region. No signature of MnP 
was observed in the photoemission spectra.

We thank T. Okuda, A. Harasawa, and T. Kinoshita for their valuable 
technical help, V.G. Voevodin, K. Ono, T. Komatsubara, and M. Okusawa for 
sample measurements, and T. Mizokawa 
for useful discussions and suggestions. This work was supported by 
a Grant-in-Aid for Scientific Research in Priority Area ``Semiconductor 
Spintronics'' (No.\ 14076209) from the Ministry of Education, Culture, 
Sports, Science and Technology, Japan. 
DDS thanks the University of Tokyo for hospitality 
during a part of this work. 
The experiment at Photon Factory was 
approved by the Photon Factory Program Advisory Committee 
(Proposal No.\ 01U005).

\end{document}